# Cluster Analysis of Educational Data: an Example of Quantitative Study on the answers to an Open-Ended Questionnaire

Onofrio Rosario Battaglia (*), Benedetto Di Paola (**) and Claudio Fazio (*)

(*) UOP_PERG (University of Palermo Physics Education Research Group)
Dipartimento di Fisica e Chimica, Università di Palermo, Italia

(**) GRIM (Mathematics Education Research Group)
Dipartimento di Matematica e Informatica, Università di Palermo, Italia

**Abstract**

In the last years many studies examined the consistency of students' answers in a variety of contexts. Some of these papers tried to develop more detailed models of the consistency of students' reasoning, or to subdivide a sample of students into intellectually similar subgroups. The problem of taking a set of data and separating it into subgroups where the elements of each subgroup are more similar to each other than they are to elements not in the subgroup has been extensively studied through the methods of Cluster Analysis. This method can separate students into groups that can be recognized and characterized by common traits in their answers, without any prior knowledge of what form those groups would take (unbiased classification). In this paper we start from a detailed analysis of the data coding needed in Cluster Analysis, in order to discuss the meaning and the limits of the interpretation of quantitative results. Then two methods commonly used in Cluster Analysis are described and the variables and parameters involved are outlined and criticized. Section III deals with the application of these methods to the analysis of data from an open-ended questionnaire administered to a sample of university students, and the quantitative results are discussed. Finally, the quantitative results are related to student answers and compared with previous results reported in the literature, by pointing out the new insights resulting from the application of such new methods.

## I. Introduction

Extensive qualitative research involving open answer questionnaires has provided instructors/teachers with tools to investigate their students' conceptual knowledge of various fields of physics. Many of these studies examined the consistency of students' answers in a variety of situations. Others looked at problems where the underlying physical systems are similar from the point of view of an expert. In recent years, some papers tried to develop more detailed models of the consistency of students' reasoning, or to subdivide a sample of students into intellectually similar subgroups. Bao and Redish [1] introduced model analysis as a framework for exploring the structure of the consistency of the application of student knowledge, by separating a group of students into intellectually similar subgroups.

The problem of taking a set of data and separating it into subgroups where the elements of each subgroup are more similar to each other than they are to elements not in the subgroup has been extensively studied through the methods of Cluster Analysis *(ClA)*. *ClA* can separate students into groups that can be recognized and characterized by common traits in their answers, without any prior knowledge of what form those groups would take (unbiased classification).

*ClA*, introduced in Psychology by R.C. Tyron in 1939 [2], has been the subject of research since the beginning of the 1960s, with its first systematic use by Sokal e Sneath [3] in 1963. The application of techniques related to *ClA* is common in many fields, including information technology, biology, medicine, archeology, econophysics and market research [4, 5, 6, 7]. For example, in market research it is important to classify the key elements of the decision-making processes of business strategies as the characteristics, needs and behavior of buyers. These techniques allow the researcher to locate subsets or clusters within a set of objects of any nature that have a tendency to be homogeneous "in some sense". The results of the analysis should reveal a high homogeneity within each group (intra-cluster), and high heterogeneity between groups (inter-clusters), in line with the chosen criteria.

*ClA* techniques [8] are exploratory and do not necessarily require a-priori assumptions about the data, but they do need actions and decisions to be taken before, during and after analysis. The selection of variables, the choice of the criteria of similarity between the data, the choice of clustering techniques, the selection of the number of groups to be obtained and the evaluation of the solution found, as well as the choice between possible alternative solutions, are particularly important. It is also important to bear in mind that different choices can lead to separate, and somehow arbitrary, results (as they heavily depend on the criteria used for the selection of the data). Subjectivity is common to all multivariate analysis methods, and is typical of the processes of reduction and controlled simplification of information.

In the literature concerning research in education, some studies using *ClA* methods are found. They group and characterize students' responses by using open-ended questionnaires [9, 10, 11] or multiple-choice tests [12]. All these papers show that the use of cluster analysis leads to identifiable groups of students that make sense to researchers and are consistent with previous results obtained using more traditional methods. Particularly, the groups of responses in open-answer questions about two-dimensional kinematics [9] identified by cluster analysis methods show striking similarity to response patterns previously reported in the literature and also provide additional information about unexpected differences between groups. In papers [10] and [11], students' responses to specially designed written questionnaires are quantitatively analyzed using researcher-generated categories of reasoning, based on the physics education research literature on student understanding of relevant physics content. Through cluster analysis methods groups of students showing remarkable similarity in the reasoning categories are identified and the consistency of their deployed mental models is validated by comparison with researcher-built ideal profiles of student behavior known from previous research. Paper [12] analyzes five commonly used approaches to analyzing multiple-choice test data (classic test theory, factor analysis, cluster analysis, item response theory and model analysis) and shows that cluster analysis is a good method to point out how student response patterns differ so as to classify students.

A recent paper [13] analyzes the evolution of student responses to seven contextually different versions of two Force Concept Inventory questions, by using a model analysis for the state of student knowledge and *ClA* methods to characterize the distribution of students' answers. This paper shows that *ClA* methods are an effective way to examine the structure of student understanding and can produce subgroups of a data sample mathematically well defined and meaningful for the researcher. The authors conclude that *ClA* is an effective method to extract the underlying subgroups in student data and that additional insight may be gained from a careful, qualitative analysis of clustering results. In fact, each cluster is characterized by means of a careful reading of the typical trends in the answers of the individuals that are part of the cluster.

It is well known that there are inherent difficulties in the classification of student responses in the studies mainly involving open-ended questionnaires. In fact, the problem of quantifying qualitative data has been widely discussed in the literature for many years [14], and it has been pointed out that, very often, a small or even unconscious researcher bias means that the categories picked out tend to find those groups of students that the researcher is already looking for. A recent paper [15] points out that researchers "*should not treat coding results as data but rather as tabulations of claims about data and that it is important to discuss the rates and substance of disagreements among coders*" and proposes guidelines for the presentation of research that quantifies qualitative data. Another paper [16] discussed the need to describe the process of developing a coding scheme, by outlining that in the process of quantifying qualitative data, the term "data" means the qualitative records supplied by students and not the result of the coding scheme.

*ClA* can be carried out using various algorithms that significantly differ in their notion of what constitutes a cluster and how to effectively find them. Notions of clusters include groups with small distances between the cluster members, dense areas of data space, intervals or particular statistical distributions. The appropriate clustering algorithms and parameter settings depend on the individual data set and intended use of the results. Moreover, a deep analysis of the *ClA* procedures applied is needed, because they often include approximations strongly influencing the interpretation of results.

For these reasons, we start from a detailed analysis of the data coding needed in *ClA*, in order to discuss the meanings and the limits of the interpretation of quantitative results. Then, two methods commonly used in *ClA* are described and the variables and parameters involved are outlined and criticized. Section III deals with the application of these methods to the analysis of data from an open-ended questionnaire administered to a sample of university students, and discusses the quantitative results. In the last section we analyze the results of *ClA* procedures in order to give meaning to quantitative results and compare them with previous results reported in the literature, by pointing out the new insights resulting from the application of such *ClA* methods.

## II. Quantitative Analysis

### A. Data setting

Research in education that uses open-ended questions and is aimed at quantifying qualitative data usually involves the development of coding procedures. This requires an analysis of student answers in order to reveal (and then examine) patterns and trends, and to find common themes emerging from them. These themes are, then, developed and grouped in a number of categories, which can be considered the typical "answering strategies" put into action by the *N* students tackling the questionnaire. Therefore, it is possible to summarize the whole set of answers given to the questionnaire into a limited number, *M*, of answering strategies, making the subsequent analysis easier. Some details are supplied in Section III.C. So, through coding and categorization, each student, *i*, can be identified by an array, $a_i$, composed of *M* components 1 and 0, where 1 means that the student used a given answering strategy to respond to a question and 0 means that he/she did not use it. Then, a *M* x *N* binary matrix (the "matrix of answers") modeled on the one shown in Table I, is built. The columns in it show the *N* student arrays, $a_i$, and the rows represent the *M* components of each array, i.e. the *M* answering strategies.

Table I
Matrix of data: the *N* students are indicated as $S_1$, $S_2$, …,$S_N$., and the *M* answering strategies as $AS_1$, $AS_2$, , $AS_M$.

| *Strategy* | *Student* | | | |
|---|---|---|---|---|
| | $S_1$ | $S_2$ | … | $S_N$ |
| $AS_1$ | 1 | 0 | … | 0 |
| $AS_2$ | 1 | 0 | … | 1 |
| … | 0 | … | … | … |
| $AS_5$ | 1 | … | … | |
| … | 0 | … | … | … |
| $AS_M$ | 0 | 1 | … | 0 |

For example, let us say that student $S_1$ used answering strategies $AS_1$, $AS_2$ and $AS_5$ to respond to the questionnaire questions. Therefore, column $S_1$ in Table I will contain the binary digit 1 in the three cells corresponding to these strategies, while all the other cells will be filled with 0.

The matrix depicted in Table I contains all the information needed to describe the sample behavior with respect to the questionnaire items. However, it needs some elaboration in order to make this information understandable. *ClA* classifies subset behaviors in different groups (the clusters). These groups can be analyzed in order to deduce their distinctive characteristics and point out similarities and differences among them.

### B. Distance

*ClA* requires the definition of new quantities that are used to build the grouping, like the "similarity" or "distance" indexes. These indexes are defined by starting from the *M* x *N* binary matrix discussed above.

In the literature [2,6,8] the similarity between two students *i* and *j* of the sample is often expressed by taking into account the distance, $d_{ij}$, between them (which actually expresses their "dissimilarity", in the sense that a higher value of distance involves a lower similarity).

A distance index can be defined by starting from the Pearson's correlation coefficient. It allows the researcher to study the correlation between students *i* and *j* if the related variables describing them are numerical. If these variables are non-numerical variables (as in our case, where we are dealing with the arrays $a_i$ and $a_j$ containing a binary symbolic coding of the answers of students *i* and *j*, respectively), we propose a modified form of the Pearson's correlation coefficient, $R_{mod}$, similar to that defined by Tumminello et al. [17]. We define $R_{mod}$ as,

$$R_{\text{mod}}(a_i, a_j) = \frac{p(a_i \cap a_j) - \dfrac{p(a_i) p(a_j)}{M}}{\sqrt{p(a_i) p(a_j) \left(\dfrac{M - p(a_i)}{M}\right)\left(\dfrac{M - p(a_j)}{M}\right)}} \qquad (1)$$

Where $p(a_i)$, $p(a_j)$ are the number of properties of $a_i$ and $a_j$ explicitly present in our students (i.e. the numbers of 1's in the arrays $a_i$ and $a_j$, respectively), *M* is the total number of properties to study (in our case, the answering strategies) and $p(a_i \cap a_j)$ is the number of properties common to both students *i* and *j* (the common number of 1's in the arrays $a_i$ and $a_j$). $\left[p(a_i)p(a_j)\right]/M$ is the expected value of the properties common to $a_i$ and $a_j$.

By following eq. (1) it is possible to find for each student, $i$, the $N-1$ correlation coefficients $R_{mod}$ between him/her and the others students (and the correlation coefficient with him/herself, that is, clearly, 1). All these correlation coefficients can be placed in a $N$x$N$ matrix that contains the information we need to discuss the mutual relationships between our students.

The similarity between students $i$ and $j$ can be defined by choosing a type of metric to calculate the distance $d_{ij}$. Such a choice is often complex and depends on many factors. If we want two students, represented by arrays $a_i$ and $a_j$ and negatively correlated, to be more dissimilar than two positively correlated (as is often advisable in research in education), a possible definition of the distance between $a_i$ and $a_j$, making use of the modified correlation coefficient, $R_{mod}(a_i, a_j)$, is:

$$d_{ij} = \sqrt{2\left(1 - R_{mod}\left(a_i, a_j\right)\right)} \qquad (2)$$

This function defines an Euclidean metric [18], which is required for the following calculations. A distance $d_{ij}$ between two students equal to zero means that they are completely similar, while a distance $d_{ij} = 2$ shows that the students are completely dissimilar. By following eq. (2) we can, then build a new $N$x$N$ matrix, D, containing all the mutual distances between the students. The main diagonal of D is composed by 0s (the distance between a student and him/herself is zero). Moreover, D is symmetrical with respect to the main diagonal.

### C. Clustering techniques: a general view

Clustering Analysis methods can be roughly distinguished in *Non-Hierarchical* (or *Centroid-Based*), and *Hierarchical* ones (also known as *connectivity based clustering* methods). The first category of methods basically takes to partitions of the data space into a structure known as a *Voronoi Diagram* (a number of regions including subsets of similar data). The second one is based on the core idea of building a binary tree of the data that are then merged into similar groups. This tree is a useful summary of the data that are connected to form clusters based on their known distance, and it is sometimes referred to as a *dendrogram*.

*Non-Hierarchical Clustering Analysis (NH-ClA)*
Non-hierarchical clustering analysis is used to generate groupings of a sample of elements (in our case, students) by partitioning it and producing a smaller set of non-overlapping clusters with no hierarchical relationships between them. Among the currently used *NH-ClA* algorithm, we will consider the *k-means* one, which was first proposed by MacQueen in 1963 [19].

The starting point is the choice of the number, $q$, of clusters one wants to populate and of an equal number of "seed points", randomly selected in the bi-dimensional Cartesian plane representing the data. The students are then grouped on the basis of the minimum distance between them and the seed points. Starting from an initial classification, students are transferred from one cluster to another or swapped with students from other clusters, until no further improvement can be made. The students belonging to a given cluster are used to find a new point, representing the average position of their spatial distribution. This is done for each cluster $Cl_k$ ($k = 1, 2, ..., q$) and the resulting points are called the cluster *centroids* $C_k$. This process is repeated and ends when the new centroids coincide with the old ones. As we said above, the spatial distribution of the set elements is represented in a 2-dimensional Cartesian space, creating what is known as the k-means graph (see Figure 1).

*NH-ClA* has some points of weakness and here we will describe how it is possible to overcome them. The first involves the a-priori choice of the initial positions of the centroids. This is usually resolved in the literature [13] by repeating the clustering procedure for several values of the initial conditions and selecting those that lead to the minimum values of the distances between each centroid and the cluster elements. Furthermore, at the beginning of the procedure, it is necessary to arbitrarily define the number, $q$, of clusters. A method widely used to decide if this number $q$, initially used to start the calculations, is the one that best fits the sample element distribution is the calculation of the so-called *Silhouette Function, S,* [20,21]. The *S* values allow us to decide if the partition of our sample in $q$ clusters is adequate, how dense a cluster is, and how well it is differentiated from the others.

Several values of the function S are calculated once a value of the number of clusters, $q$, is fixed:
- the individual value, $S_i(q)$ for each student, $i$, of the sample. It gives a measure of how similar student $i$ is to the other students in its own cluster $Cl_k$, when compared to students in other clusters. It ranges from -1 to +1; a value near +1 indicates that student $i$ is well-matched to its own cluster, and poorly-matched to neighboring clusters. If most students have a high silhouette value, then the clustering solution is appropriate. If many students have a low or negative silhouette value, then the clustering solution could have either too many or too few clusters (i.e. the chosen number, $q$, of clusters should be modified).
- The average silhouette value in cluster $Cl_k$, $\langle S(q) \rangle_k$, with $k=1, 2,...q$. It gives the average value of $S_i(q)$, calculated on all the students belonging to cluster $Cl_k$ and it is a measure of the density of the cluster. Large values of $\langle S(q) \rangle_k$ (close to 1) are to be related to cluster elements being tightly arranged in the cluster $k$, and vice versa [20,21].
- The total average silhouette value, $\langle S(q) \rangle$ for the chosen partition in $q$ clusters. It gives the average value of $S_i(q)$, calculated on all the students belonging to the sample. Large values of $\langle S(q) \rangle$ are to be related to well defined clusters [20,21]. It is, therefore, possible to perform several repetitions of the cluster calculations (with different values of $q$) and to choose the number of clusters, $q$, that gives the maximum value of $\langle S(q) \rangle$.

The k-means results can be plotted in a 2-dimensional Cartesian space containing points that represent the students of the sample placed in the plane according to their mutual distances. As we said before, for each student, $i$, we know the $N$ distances, $d_{ij}$ between such a student and all the students of the sample (being $d_{ii}$ = 0 ). It is, then, necessary to define a procedure to find two Cartesian coordinates for each student, starting from these $N$ distances. This procedure consists in a linear transformation between a $N$-dimensional vector space and a 2-dimensional one and it is well known in the specialized literature as *multidimensional scaling* [22].

Figure 1 shows an example of the spatial distribution of the results of a k-means analysis on a same set of data, represented in a 2-dimensional Cartesian space. First three clusters ($q = 3$ in Fig. 1a), and then four ($q = 4$ in Fig. 1b) have been chosen to start the calculations. The x- and y-axes simply report the values needed to place the points according to their mutual distance. In this specific case, $\langle S(3) \rangle > \langle S(4) \rangle$, showing that in the case of three clusters these are more defined than in the other one. In both cases $\langle S(3) \rangle_1$ and $\langle S(4) \rangle_1$ have the maximum value among the $\langle S(q) \rangle_k$ ones, showing that cluster $Cl_1$ is denser, and more compact than the other ones. A more detailed discussion on the Silhouette function can be found in Appendix A.1

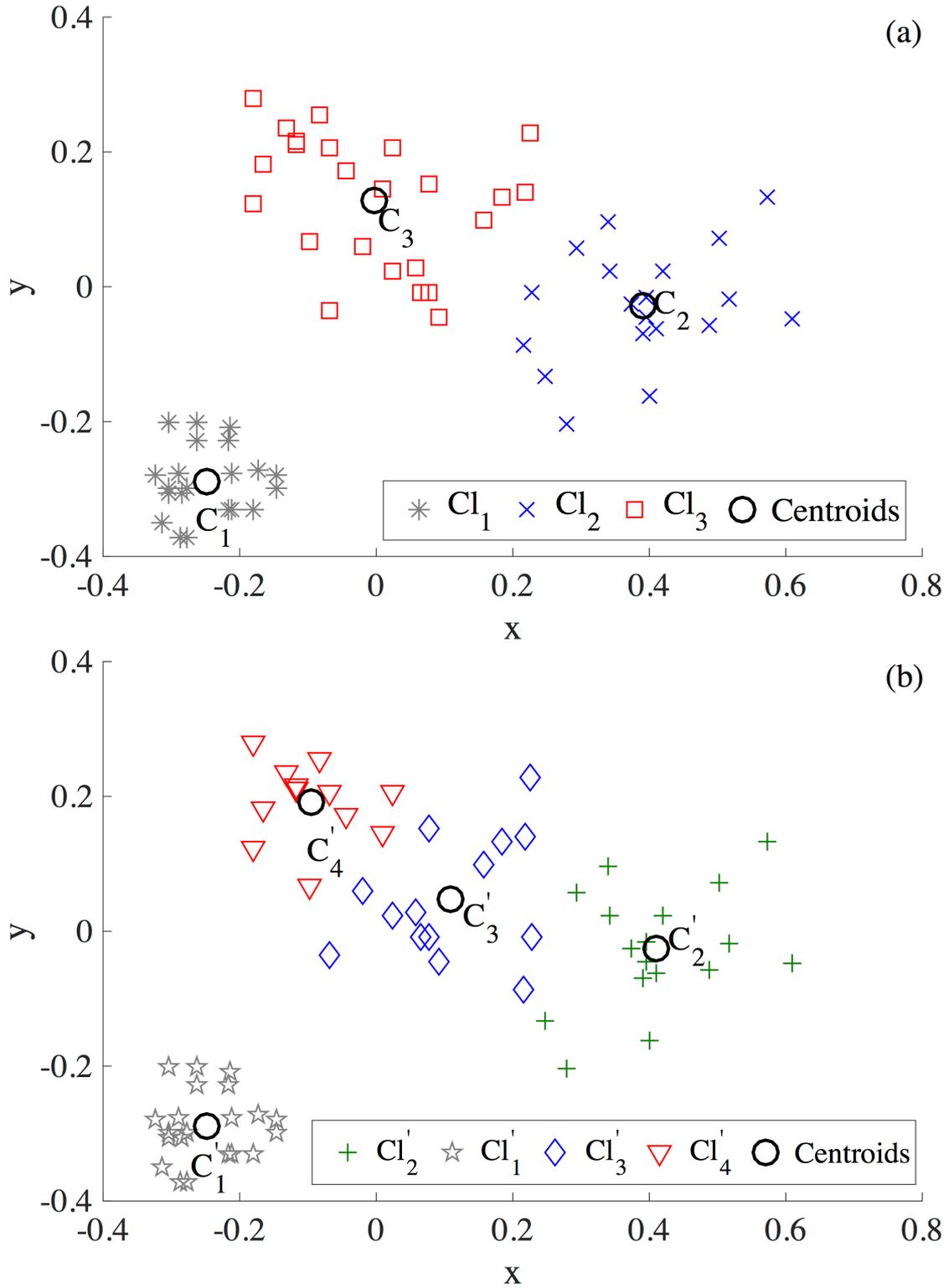

**Figure 1:** Clustering of $N$=64 data using k-means method. Fig1a) shows $q = 3$ clusters with $\langle S(3)\rangle$= 0.795, with a 95% confidence interval[1], CI, equal to (0.780, 0.805) and $\langle S(3)\rangle_1$= 0.953, CI = (0.951, 0.956) , $\langle S(3)\rangle_2$=0.79, CI = (0.78, 0.81), and $\langle S(3)\rangle_3$ =0.66, CI = (0.63, 0.68). Fig1b) shows $q = 4$ clusters with $\langle S(4)\rangle$ = 0.729, CI = (0.711, 0.744), and $\langle S(4)\rangle_1$ = 0.953, CI = (0.951, 0.956), $\langle S(4)\rangle_2$ = 0.67, CI = (0.64, 0.69), $\langle S(4)\rangle_3$ =0.77, CI = (0.74, 0.79) and $\langle S(4)\rangle_4$ = 0.44, CI = (0.40, 0.47).

---

[1] The confidence intervals have been calculated by using the Bootstrap method [23,24] as the distribution of the Silhouette values are not a-priori known.

It is interesting to study how well a centroid geometrically characterizes its cluster. Two parameters affect this: both the cluster density and the number of its elements.[2] For this purpose, we propose a coefficient, $r_k$, defined as the centroid *reliability*. It is calculated as follows:

$$r_k = \frac{\langle S(q) \rangle_k}{1 - \langle S(q) \rangle_k} \frac{1}{n_k} \qquad (3)$$

where $n_k$ is the number of students contained in cluster $Cl_k$ and $\langle S(q) \rangle_k$ is the average value of the *S-function* on the same cluster, that, as we pointed out, gives information on the cluster density.[3] High values of $r_k$ indicate that the centroid characterizes the cluster well.

In order to compare the reliability values of different clusters in a given partition the $r_k$ values can be normalized according to the following formula

$$r_k^{norm} = \frac{r_k - \langle r_k \rangle}{\sigma(r_k)}$$

where $\langle r_k \rangle$ and $\sigma(r_k)$ are the mean value and the variance of $r_k$ on the different clusters, respectively.

Once the appropriate partition of data has been found, we want to characterize each cluster in terms of the most prominent answering strategies. Such characterizations will help us to compare clusters and relate our findings to the literature. To do this, we start by creating a 'virtual student' for each of the *q* clusters, $Cl_k$ (with $k = 1,2,..., q$), represented by the related centroids. Since each real student is characterized by an array $a_i$ composed by 0 and 1 values for each of the *M* answering strategies, the array for the virtual student, $\bar{a}_k$, should also contain *M* entries with 0's for strategies that do not characterize $Cl_k$ and 1 for strategies that do characterize $Cl_k$. It is possible to demonstrate that $\bar{a}_k$ contains 1 values exactly in correspondence to the answering strategies most frequently used by students belonging to $Cl_k$.[4] In fact, since a centroid is defined as the geometric point that minimizes the sum of the distances between it and all the cluster elements, by minimizing this sum the correlation coefficients between the cluster elements and the virtual student are maximized and this happens when each virtual student has the largest number of common strategies with all the students that are part of its cluster. This is a remarkable feature of $\bar{a}_k$, that validates our idea to use it to characterize cluster $Cl_k$.[5]

*Hierarchical Clustering Analysis (H-ClA)*
Hierarchical clustering is a method of cluster construction that starts from the idea of elements (again students in our case) of a set being more related to nearby students than to farther away ones, and tries to arrange students representing them as being "above", "below", or "at the same level as" one another. This method connects students to form clusters based on the presence of common characteristics. As a *hierarchy* of clusters, which merge with each other at certain distances, is provided, the term "hierarchical clustering" has been used in the literature.

In *H-ClA*, which is sometimes used in education to analyze the answers given by students to open- and closed-ended questionnaires (see, for example, [9, 11, 12]), each student is initially considered as a separate cluster. Then the two 'closest' students are joined as a cluster and this process is continued (in a stepwise manner) to join one student with another, a student with a cluster, or a cluster with another cluster, until all the students are combined into one single cluster as one moves up the hierarchy (*Agglomerative Hierarchical*

---

[2] For example, two clusters with similar density but different numerosity (i.e. with different variability of student properties) are differently characterized by their centroids: the more populated cluster being less well characterized by its centroid than the other one.

[3] The term $1 - \langle S(q) \rangle_k$ in (3) is needed to differently weight $\langle S(q) \rangle_k$ and $n_k$ because when $\langle S(q) \rangle_k \to 1$ the $r_k$ value should be high independently off the value of $n_k$.

[4] It is worth noting that if some answering strategies are only slightly more frequent than the other ones all those with similar frequencies should also be considered.

[5] Appendix A.2 reports and alternative method to obtain the array $\bar{a}_k$ based on an iterative method.

*Clustering*). Another possibility is to build recursive partitions from a single starting cluster that contains all the students observed (*Divisive Hierarchical Clustering*). The results of hierarchical clustering are graphically displayed as a tree, referred to as the *hierarchical tree* or *dendrogram*. The term 'closest' is identified by a specific rule in each of the *linkage methods*. Hence, in different linkage methods, the corresponding distance matrix after each merger are differently computed.

Among the many linkage methods described in the literature, the following have been taken into account in education studies: *Single, Complete, Average and Weighted Average*. Each method differs in how it measures the distance between two clusters *r* and *s* by means of the definition of a new metric (an 'ultrametric'), and consequently influences the interpretation of the word 'closest'. *Single Linkage*, also called *nearest neighbor linkage*, links *r* and s by using the smallest distance between the students in *r* and those in *s*; *Complete Linkage*, also called *farthest neighbor linkage*, uses the largest distance between the students in *r* and the ones in *s*; *Average Linkage* uses the average distance between the students in the two clusters; *Weighted Average Linkage* uses a recursive definition for the distance between two clusters. If cluster *r* was created by combining clusters *p* and *q*, the distance between *r* and another cluster *s* is defined as the average of the distance between *p* and *s* and the distance between *q* and *s*.

In Appendix A.3 we present the details of the calculations for the different linkage methods. Here, to better represent the differences and approximations involved in the various linkages, an example is displayed in Figure 2 for the simple case of a sample made of 3 students. It is worth noting that in the case of clusters made up of few single students, the graph obtained by using the weighted average linkage is the same as the one obtained using average linkage. So, Figure 2 does not report the graph obtained by using the Weighted Average linkage method. A case where the difference between the two methods is more evident is reported in Appendix A.3.

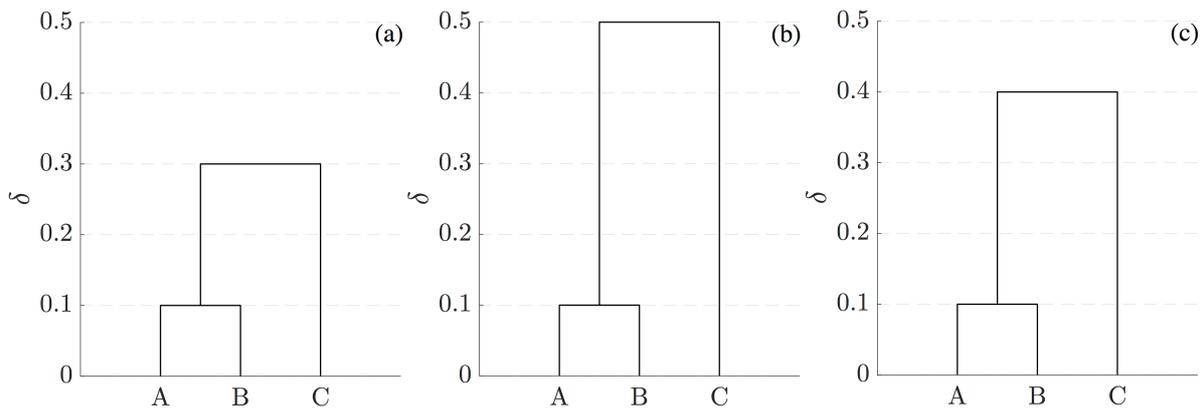

**Figure 2.** Let us suppose that for three students, A, B, and C of a set, the distances (calculated, for example, according to equation (2)) are $d_{AB} = 0.1$, $d_{AC} = 0.3$ and $d_{BC} = 0.5$. Fig. 2 (a) depicts the case of *single linkage*. Fig. 2 (b) depicts the case of *complete linkage*. Fig. 2 (c) depicts the case of *average linkage*. The δ value on the y-axis is the 'ultrametric' distance, which indicates the values of the different links between the students.

Several conditions can determine the choice of a specific linkage method. For instance, when the source data are in binary form (as in our case) the single and complete linkage methods do not give a smooth progression of the distances [9]. For this reason, when the source data are in binary form, the viable linkage methods actually reduce to the average or weighted average ones.

In the specialized literature it is easy to find numerical indexes driving the choice of a specific linkage method, such as the "*Cophenetic correlation coefficient*" [25,26]. Its value is based on the correlation (like the Pearson one [27]) between the original distances, in D, and the ultrametric distances given by the linkage type (contained in a new matrix, *Δ*), and it evaluates how much the latter are actually representative of the former. More precisely, the cophenetic coefficient is a measure of how faithfully a dendrogram preserves the pair wise distances between the original un-modeled data points (see Appendix A.4 for details). In the cases we analyzed the highest values of the cophenetic coefficient are always obtained by using average or weighted average linkage methods.

Reading a dendrogram and finding clusters in it can be a rather arbitrary process. There is not a widely accepted criterion that can be applied in order to determine the distance values to be chosen for identifying the clusters. Different criteria, named *stopping criteria*, aimed at finding the optimal number of clusters are discussed in the literature (see, for example, [27]). Many of these cannot be applied to non numerical data, as it is our case. Here we discuss two criteria applicable to our case: the first one involves the calculation of the "*Inconsistency Coefficient*" [28] and the second is known as "*Variation Ratio Criterion*" [29].

One way to decide if the grouping in a data set is adequate is to compare the height of each link in a cluster tree with the heights of neighboring links below it in the tree. A link that is approximately the same height as the links below it indicates that there are no distinct divisions between the objects joined at this level of the hierarchy. These links are said to exhibit a high level of consistency, because the distance between the objects being joined is approximately the same as the distances between the objects they contain. On the other hand, a link whose height differs noticeably from the height of the links below it indicates that the objects joined at this level in the cluster tree are much farther apart from each other than their components were when they were joined. This link is said to be inconsistent with the links below it.

The relative consistency of each link in a hierarchical cluster tree can be quantified through the Inconsistency Coefficient, $I_k$ [28] (see appendix A.5 for details). The higher is the value of this coefficient, the less consistent is the link connecting the students. A link that joins distinct clusters has a high inconsistency coefficient; a link that joins indistinct clusters has a low inconsistency coefficient.

The choice of $I_k$ value to be considered significant in order to define a threshold is arbitrary and involves the choice of the significant number of clusters that can describe the whole sample. Moreover, in the specialized literature [28] the $I_k$ value of a given link is considered by also taking into account the ultrametric distance of the link, in order to avoid a too low or too high fragmentation[6] of the sample clusters. This means that, after having disregarded the links that produce a too low fragmentation, the $I_k$ of the links just below are taken into account.

The Variation Ratio Criterion (*VRC*) [29] is also used in the literature to define the clustering validity. It measures the ratio between the sum of the squares of the distances between the elements belonging to the same cluster and the sum of the squares of the distances between the elements of a given cluster and the external ones. The larger is the *VRC* value, the better is the clustering. For more detail, see Appendix A.

It is worth noting that the evaluation of the number of cluster to be consider significant for an education-focused research should also be influenced by pedagogic considerations, related to the interpretation of the clusters that are formed. Although it could be desirable to have a fine grain description of our sample students, this can make the search for common trends in the sample too complicated, and perhaps less interesting if too many "micro-behaviors" related to the various small clusters are found and have to be explained.

As a final consideration, we want to point out that the comparison of different clustering methods (in our case *NH-ClA* and *H-ClA* methods) is a relevant point. As Meila et al. [30] point out: "*just as one cannot define a best clustering method out of the content, one cannot define a criterion for comparing clusters that fits every problem.*". Many coefficients have been identified to compare two partitions of the same data set obtained with different methods, but the majority of them are not applicable to our data that are in binary form. However a criterion, called *Variation of Information (VI),* can be applied in our case. It measures the difference in information shared between two particular partitions of data and the total information content of the two partitions. In this sense, the smaller the distance between the two clustering the more these are coherent with each other, and vice versa. *VI* can be normalized to 0-1 range: a value equal to 0 indicates very similar clustering results, and a value equal to 1 corresponds to very different ones. Reference [30] supplies all the details for *VI* calculation as well as examples of its application.

### III. Example of quantitative study

In this section we present an application of the described *ClA* procedures to the analysis of data from an open-ended questionnaire administered to a sample of university students, and discuss the quantitative results by outlining the limits of the different quantitative methods we used. The open-ended questionnaire investigates the student understanding of the modeling concept and of its main characteristics.

---

[6] A "too low" fragmentation is here to be intended as a situation where one or two big clusters are produced, that do not allow us to effectively describe the sample behavior. A "too high" fragmentation means that many small clusters, containing only a few students, are produced.

## A. The research problem

To achieve an agreed definition of model is intrinsically problematic, since this issue can be tackled under very different viewpoints, shared, for example, by psychologists, scientists and philosophers of science. However, some aspects are considered relevant in all these fields. These are in our opinion found in the works by the philosopher of science Bunge [31]. According to his viewpoint, the essential characteristics of a scientific model can be summarized in the following statement that makes explicit the ontological components, as well as the functions of a scientific model: *A scientific model is a representation of a real or conjectured system, consisting in a set of objects with its outstanding properties listed, and a set of law statements that declare the behaviors of these objects, the essential functions of a scientific model being predictions and explanations.*

The fundamental role played by models and modeling activities in the teaching/learning process in math and science education is widely recognized, and many studies present operative approaches in this direction. In particular, the conceptual "framework" presented in the report of the Committee on Conceptual Framework for New K-12 Science Education Standards [32] articulates the major practices that scientists employ in developing and using models. Such description also supplies an operational definition of what can be considered an "expert" point of view of the scientific model.

Moreover, an extended body of knowledge in the field of science education research about models involved the understanding/conceptions of models' nature of students at different school levels [11, 33, 34, 35] as well as of teachers [36, 37, 38, 39]. Such research makes explicit student epistemic criteria for evaluating scientific models and the student/teacher knowledge about scientific model, i.e. the role of the scientific model in the process of construction of knowledge, its components (that is the set of entities and laws that relate them) and its functions (that is prediction, explanation, testing) [39]. The various researchers point out different characterizations of student/teacher concept of scientific model. . In particular, Grosslight et al. [33] reported results of clinical interviews developed to elicit high school students' conceptions of models and their use in science. The paper deeply analyze how different general levels of understanding models reflect different student epistemological viewpoints and compare these with expert viewpoints. They identify three general levels of thinking about models that differ in how the relationship of model to reality is described. Level 1 involves conceptions of models that are basically consistent with a naive realist epistemology (models as physical copies of reality that embody different spatiotemporal perspectives). Level 2 modelers see models as representations of real-world objects or events and not as representations of ideas about real-world objects or events. They also see the use of different models as that of capturing different spatiotemporal views of the object rather than different theoretical views. Level 3 modelers recognize that models are constructed in the service of developing and testing ideas (rather than as serving as a copy of reality itself) and can be manipulated and subjected to tests in the service of informing ideas.

Treagust et al. [34] discuss the development of an instrument to measure secondary students' understanding of scientific models and report the results of a study with a sample of secondary science students. They identify five factors about students' understanding of scientific models: scientific models as multiple representations; models as exact replicas; models as explanatory tools; how scientific models are used; and the changing nature of scientific models. Their results supply percentages of student answers for the different factors and, although such percentages are obviously different from those reported in previous studies (for the difference in the contexts), their conclusions can be considered consistent with them.

Pluta et al. [35] analyze epistemic criteria for good scientific models generated by a sample of middle-school students, after evaluating a range of models, but before extensive instruction or experience with model-based reasoning practices. In this study, students, following the questions posed by teachers, generated lists of criteria of good scientific models. Students' individual lists of criteria were compared to expert criteria (identified by philosophers of science, and with findings from previous research on students' understanding of modeling). Through a deep analysis of such lists the authors show that their students seemed to have a wide range of ideas about one important element of the epistemic practices of science-the epistemic criteria for good models. In fact they outline that many of the criteria proposed by students are similar to the criteria used by scientists (as identified by philosophers of science). Primary epistemic criteria include criteria related to the explanatory function of models, the role of evidence, appropriate details, and accuracy. Moreover, the presence in the lists of criteria related to the communicative or constituent features of models suggest to the authors that students see models as more than just direct copies or scale models. The authors recognize that their results differ from some of previous research on students' naïve understanding of both the nature of science and modeling practice and outline that this can be ascribed to the different context of the research as well as to the major focus posed by teacher questions to modeling procedures and functions of scientific models.

Some studies involving teacher conception of scientific models [38, 39] report conceptions of models defined as "realistic" [39] since they involve exact/partial replica of reality. Only few show a knowledge about scientific models fitting expert conceptions and some time such a knowledge is incomplete or declarative and showing its limits when more information are required about functions and or characteristics of scientific models . Incoherent responses resulting in a very poor understanding of the scientific models are pointed out mainly when teachers focus on the role of models as examples of objects/processes or their simplifications.

It is also noteworthy to analyze researches [11, 40] concerning the characterization of students' reasoning in different physics fields aimed at studying how students use models in the understanding of simple phenomena. In order to identify the kind of reasoning related to student mental models, the researchers define some kinds of reasoning as ''hybrid models'' [12] or ''synthetic models'' [37], by referring to composite mental models that unify different features of initial spontaneous models and scientifically accepted ones. Research reveals [1, 40] that a student can use different mental models in response to a set of situations or problems considered equivalent by an expert. In particular, Bao and Redish [1] developed a way to deal with these composite mental models and defined students' model states that can change with specific contextual features in different equivalent questions. We think that it is also possible that such kind of reasoning can be connected with an hybrid conception of the semantic of scientific models and their role in the construction of physics knowledge.

Our research aims at analyzing the understanding of scientific model concept by a sample of university students, through the analysis of their answers to an open answer questionnaire investigating the definition of scientific model, its main constituents and its functions. The methodological approach aims at pointing out clusters of students that share representations of scientific model making sense to the researcher. Here, "to make sense to the researcher" means that such representations present a logical coherence and/or have been already described in the literature.

Students' answers have been empirically coded and then quantitatively analyzed, as described in the next section. In particular, we used the two *ClA* methods we described in Section II, in order to identify such clusters. The specific research questions that guided our study are:

- RQ1. How are the two different *ClA* methods effective in partitioning students into groups that can be characterized by common traits in students' answers and how can the results be combined to create a coherent characterization of the data?
- RQ2. How do the common traits in students' concept of scientific model identified by *ClA* method relate to literature findings and what new insights do they supply?

**B. The questionnaire and the sample**

The questionnaire is made up of four questions, which focus on the understanding of the modeling concept (see Appendix B). They are part of a more complex questionnaire, which has already been used in previous research [10, 41]. We chose to analyze a questionnaire with a low number of questions, and consequently a relatively low number of answers, in order to be able to easily relate quantitative results to student answers.

The four selected items refer to: I) the definition of a physics model, II) the students' beliefs about the representational modes of physics models, III) the main characteristics of models, and IV) the students' beliefs about the modeling process.

The questionnaire was administered to 124 freshmen of the Information and Telecommunications Engineering Degree Course at the University of Palermo, during the first semester of the academic year 2013/2014. The students were given the questionnaire during the first lesson of general physics, before any discussion on the model concept had started.

**C. Categorization of student answers**

The three authors independently read the students' answers in order to empirically identify the main characteristics of the different student records (the raw data). Each author independently constructed a coding scheme consisting in the identification of keywords, which characterized student answers. During a first meeting, the selected keywords were compared and contrasted, and then grouped into categories based on epistemological and linguistic similarities[7]. These categories were also re-analyzed through the authors' interactions with the data, and taking into account the existing literature about models and modeling [33, 34, 35]. The complete list of 20 categories shared by researchers with respect to the four questions is reported in Appendix B, where examples of specific student answers are also supplied. To give an example of the categorization procedure we refer to question Q1 where the categories have been defined as follows:

- Category 1A groups all answers where the model conception reflects a confusion between model and example, or general law , or procedure, or rules or experiment  with the objective of describe phenomena.
- Student answers grouped in category 1B present a conceptions of model as a simple copy/replica of reality in the case objects (scale models) or phenomena (a simple experiment that models a phenomenon) are referred to. This category can be reported to Level 1 described in paper [33] or to theme 2 (models as exact replicas) described in paper [34] and many answers supplied by our student sample are similar to those reported in such papers [33, 34].
- Category 1C groups all answers where the model is clearly presented as a representation (pictures, mathematical expressions, diagrams…. ,  of an entity, simple or complex, that displays a particular perspective or emphasis aimed at describing (or understanding how) such entity behaves.
- Student answers grouped in category 1D present a conception of model as a representation aimed at explaining (to understand why, to explain what happens, to supply a mechanism of functioning, to provide answers to a problem, to predict behaviors,…). Such conceptions of "models as explanatory tools" [34] can be reported to characteristics of Level 3 conceptions reported in paper [33] as well as to the primary epistemic criteria discussed in paper [35].

As a third step of the categorization process, each researcher read again the student records and applied the new coding scheme, by assigning each student to a given category for each question. Given the inevitable subjectivity of the researchers' interpretations, the three lists were compared and contrasted in order to get to a single agreed list. Discordances between researcher lists were usually a consequence of different researchers' interpretations of student statements. This happened 14 times when comparing tables of researchers 1 and 2, 9 times for researchers 2 and 3, and 12 times for researchers 1 and 3. Hence we obtained very good percentages of accordance (97%, or higher) between the analysis tables of each researcher pair. When a consensus was not obtained, the student answer was classified in the category "not understandable statement".

As a result of the coding and categorization, we obtain a matrix like the one depicted in Table I, where $N = 124$ and $M = 20$. This matrix of data represents a set of properties (the categories to which student answers have been assigned) for each sample element (the student being analyzed).

**D. Data analysis**

All the clustering calculations were made using a custom software, written in C language, for the *NH-ClA* (k-means method) as well as for *H-ClA*, where the weighted average linkage method was applied. The graphical representations of clusters in both cases were obtained using the well-known software MATLAB [42].

*Non-hierarchical clustering analysis (NH-ClA)*

In order to define the number $q$ of clusters that best partitions our sample, the mean value of *S-function*, $\langle S(q) \rangle$, has been calculated for different numbers of clusters, from 2 to 10 (see Figure 3)[8]. The figure shows that the best partition of our sample is achieved by choosing four clusters, where $\langle S(q) \rangle$ has its maximum. The obtained value $\langle S(4) \rangle$ = 0.617, with a 95% confidence interval CI = (0.607, 0.625), indicates that a reasonable cluster structure has been found (see Appendix A1).

---

[7] For example, students that defined models as *simple phenomena* or *experiments* or *reproductions of an object on a small scale* have been put on the same category since the three definitions have been intended as giving a ontological reality to models.

[8] As discussed in Section II.C, for each value of *q* the clustering procedure was repeated for several values of the initial conditions. In each case, we selected the cluster solution that leads to the minimum values of the distances between each centroid and the cluster elements.

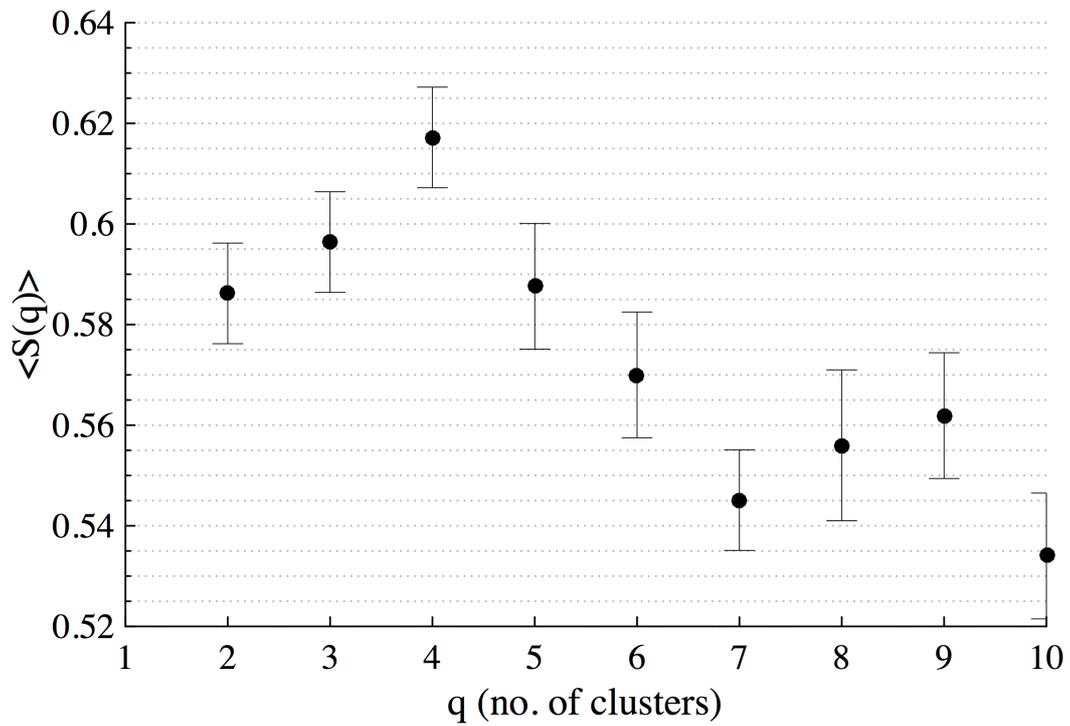

Figure 3: Average Silhouette values and related 95% confidence intervals (CI) for different cluster partitions of our sample. The two highest values are obtained for partitions in $q = 4$ clusters ($\langle S(4) \rangle = 0.617$, CI = (0.607, 0.625)) and in $q = 3$ clusters ($\langle S(3) \rangle = 0.596$, CI = (0.586, 0.603)).

Figure 4 shows the representation of this partition in a 2-dimensional graph. The four clusters show a partition of our sample into groups made up of different numbers of students (see Table II)

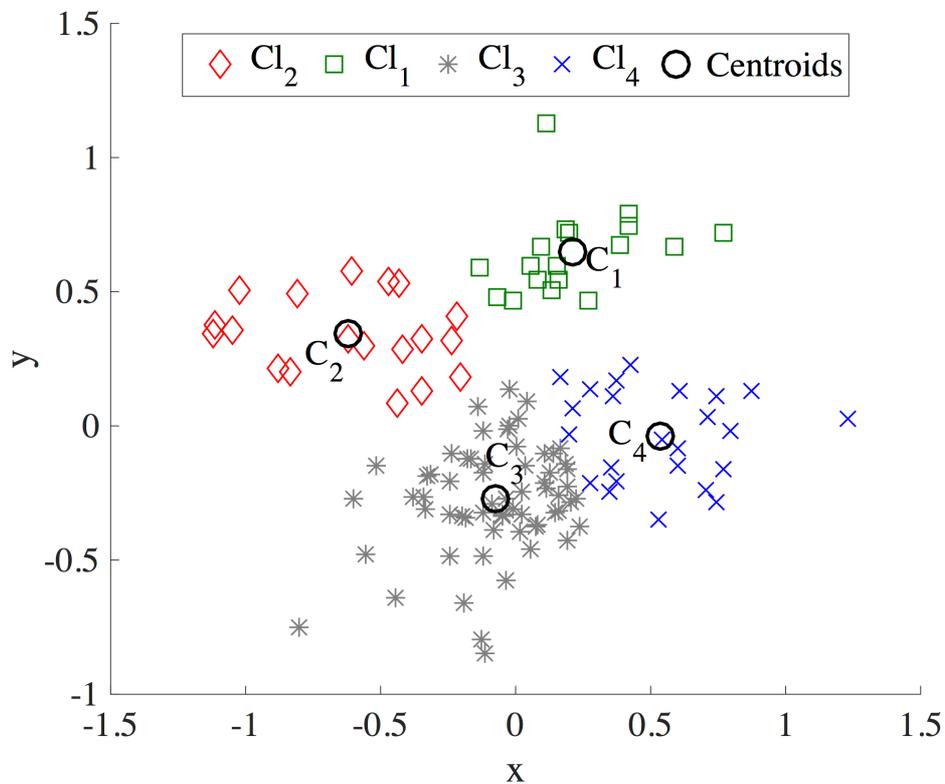

Figure 4: *K-means* graph. Each point in this Cartesian plane represents a student. Points labeled $C_1, C_2, C_3, C_4$ are the centroids.

The four clusters $Cl_k$ ($k=1,...,4$) can be characterized by their related centroids, $C_k$. They are the four points in the graph whose arrays, $\bar{a}_k$, contain the answering strategies most frequently applied by students in the related clusters (see Table II). The codes used refer to the answering strategies for the questionnaire items described in Appendix B. Table II also shows the number of students in each cluster, the mean values of the *S-function* $\langle S(4) \rangle_k$ ($k=1,..,4$) for the four clusters and the normalized reliability index $r_k^{norm}$ of their centroids.

Table II
An overview of results obtained by NH-ClA method

| Cluster centroid | $C_1$ | $C_2$ | $C_3$ | $C_4$ |
|---|---|---|---|---|
| $\bar{a}_k$ (Most frequently given answers) | 1B, 2C, 3B, 4A | 1B, 2B, 3E, 4A | 1C, 2B, 3A, 4A | 1C, 2C, 3B, 4B |
| Number of students | 18 | 19 | 63 | 24 |
| $\langle S(4) \rangle_k$ | 0.750, CI = (0.730, 0.763) | 0.62, CI = (0.58, 0.64) | 0.604, CI = (0.590, 0.616) | 0.56, CI = (0.53, 0.58) |
| $r_k^{norm}$ | 1.40 | -0.02 | -0.92 | -0.46 |

The $\langle S(4) \rangle_k$ value indicates that cluster $Cl_1$ is denser than the others, and $Cl_4$ is the most spread out. Furthermore, the values of $r_k^{norm}$ show that the centroid $C_1$ best represents its cluster, whereas $C_3$ is the centroid that less represents and characterizes its cluster.

*Hierarchical clustering analysis (H-ClA)*
In order to apply the *H-ClA* method to our data, we first had to choose what kind of linkage to use. Since we could not use simple or complete linkages (see Section II.C), we calculated the *cophenetic correlation coefficient* for the *average* and *weighted average* linkages, which gave a measure of the accordance between the distances calculated by (2) and the ultrametric distances introduced by the linkage. We obtained the values 0.61 and 0.68 for *average* and *weighted average* linkages, respectively. We chose to use the *weighted average* link and Figure 4 shows the obtained dendrogram of the nested cluster structure.

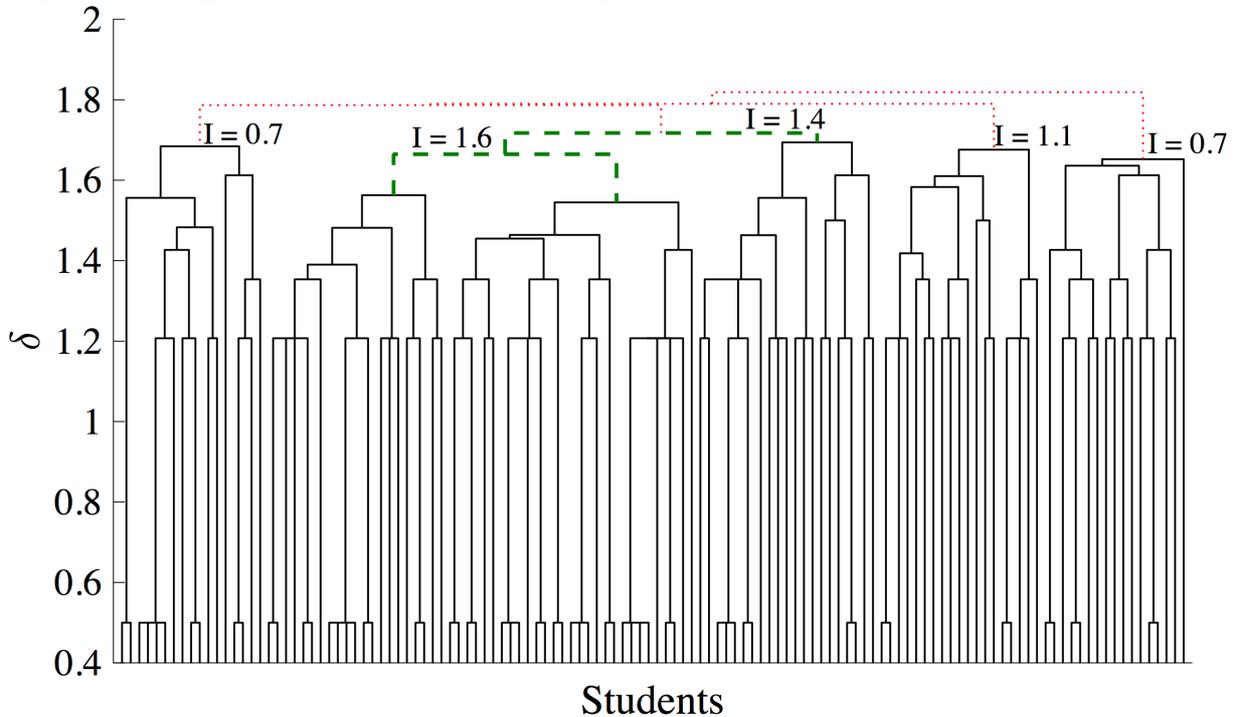

Figure 5: Dendrogram of our data. Horizontal and vertical axes represent students and ultrametric distances, respectively. Red, dotted links are at ultrametric distances of about 1.8. The Inconsistency Coefficients of the links just below these links are shown.

In this figure the vertical axis represents the ultrametric distance between two clusters when they are joined; the horizontal axis is divided in 124 ticks, each representing a student. Furthermore, vertical lines represent students or groups of students and horizontal lines represent the joining of two clusters. Vertical lines are always placed in the center of the group of students in a cluster and horizontal lines are placed at the height which corresponds to the distance between the two clusters that they join.

By describing the cluster tree from the top down, as if clusters are splitting apart, we can see that all the students come together into a single cluster, located at the top of the figure. In this cluster, for each pair of students, *i* and *j*, the ultrametric distance is $\delta_{ij} \leq 1.8$. Since the structure of the tree shows that some groups of students are more closely linked, we can identify local clusters where students are linked with distances whose values are lower than the previous one. The problem is how to find a value of distance that involves significant links. By using the *Inconsistency Coefficient*, $I_k$ (see Section II.C), we can define a specific threshold and neglect some links because they are inconsistent. In fact, this coefficient characterizes each link in a cluster tree by comparing its height with the average height of other links at the same level of the hierarchy. The choice of the threshold is arbitrary and should be limited to the links in a specific range of distances [24], yet it allows us to compare all the clusters and to treat all links with the same criterion.

If we disregard the higher links (δ ≈ 1.8, red, dotted links in Figure 5) because their use would produce a unique, single cluster of our sample, or two big ones, and we also take into account a threshold for the Inconsistency Coefficients equal to 1.6 (i.e. we consider inconsistent all the links that have $I_k > 1.6$, we can accept all the links just below, including the green, dashed ones in Figure 5 (that have $I_k$ equal to 1.4 and 1.6, respectively). So, we find a partition of our sample into 4 clusters. If, on the other hand, we introduce a lower threshold for the $I_k$ value, but still not producing a too high fragmentation, like for example $I_k > 1.25$, we must disregard the green, dashed links in the dendrogram in Figure 5, and obtain Figure 6, where 6 clusters are present. This last representation has a higher significance than the previous one since the links displayed are those that, at equal distances, show a higher consistency.

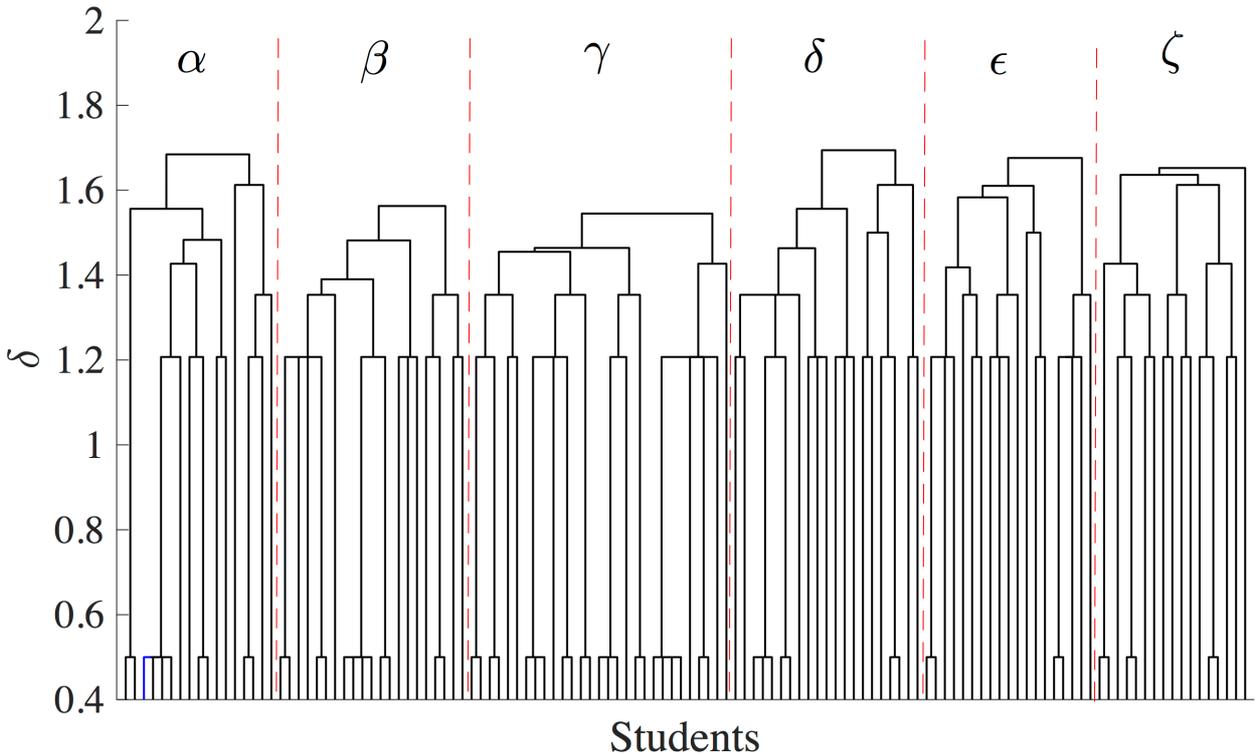

Figure 6: Dendrogram with six different clusters (α, β, γ, δ, ε, ζ), formed according to the Inconsistency Coefficient $I_k$=1.25.

Figure 6 shows the 6 distinct clusters α, β, …, ζ above identified. They can be characterized by analyzing the most frequent answers to each of the four questions in the questionnaire (see Section IV).

In order to verify the validity of our choice we also used the *VRC* (see Section II.C). Figure 7 shows the *VRC* values for different numbers of clusters. The maximum value is obtained for *q* = 6.

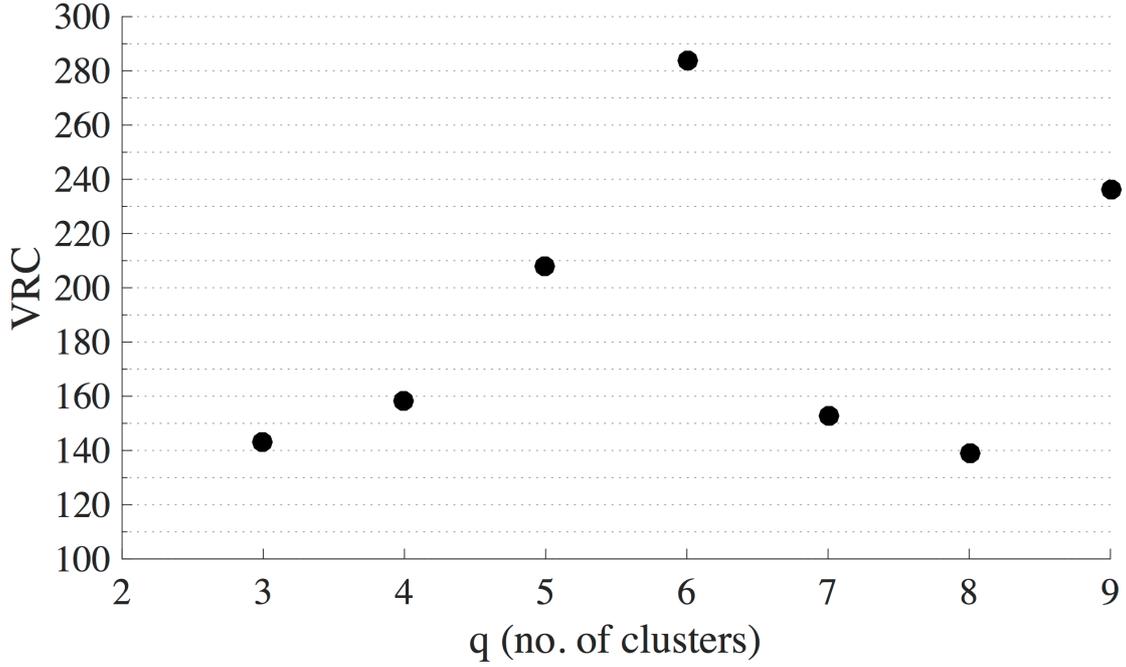

Figure 7. VRC values for some partitions of our sample in different numbers of clusters

Table III provides significant information concerning the *H-ClA* clustering.

By looking at the number of students, and at their identity, we can see that the main results of the new grouping are the redistribution of the students, originally assigned to cluster $Cl_3$ by *NH-ClA*, into different sub groups, and a redistribution of students located on the edges of cluster $Cl_4$. Furthermore, the students in cluster $Cl_1$ are all located in cluster $\beta$ and students in cluster $Cl_2$ are all located in cluster $\gamma$. This is in accordance with the high values of the $r_k^{norm}$ coefficient (shown in Table II) for $Cl_1$ and $Cl_2$ and the low value for clusters $Cl_3$ and $Cl_4$.

Table III
An overview of results obtained by *H-ClA* and comparison with those obtained by *NH-ClA* method

| Cluster | α | β | γ | δ | ε | ζ |
|---|---|---|---|---|---|---|
| Most frequently given answers | 1C, 2C, 3B, 4B | 1B, 2C, 3B, 4A/4B | 1B, 2B, 3E, 4A | 1C, 2B, 3D, 4A | 1D, 2C, 3A, 4B | 1A, 2A, 3A, 4D |
| Number of students | 17 | 21 | 29 | 21 | 19 | 17 |
| Characterization of students in cluster by the k-means method (*) | $(14)Cl_4+(3)Cl_3$ | $(18)Cl_1+(3)Cl_4$ | $(19)Cl_2+(10)Cl_3$ | $(19)Cl_3+(2)Cl_4$ | $(14)Cl_3+(5)Cl_4$ | $(17)Cl_3$ |

(*) i. e. $(14)Cl_4+(3)Cl_3$, means that cluster α contains 14 students part of the cluster $Cl_4$ (in NH-ClA) + 3 students part of cluster $Cl_3$.

In conclusion, we can say that although the two partitions of our student sample are different, they are consistent. The characterization via the dendrogram allows us to obtain greater detail. This happens in particular, in the case of cluster $Cl_3$, which turns out to be very extensive, with a large number of students and a low value of $r_k^{norm}$.

In order to better compare the results obtained by *NH-ClA* and *H-ClA* methods, we applied the variation of information (*VI*) criterion (see Section II.C), that measures the amount of information gained and lost when switching from one type of clustering to another. We calculated the value of *VI* to compare the 4-clustering results of *k-means* method with the 4-clustering, 5-clustering and 6-clustering results of *H-ClA* method and obtained the values of 0.34, 0.38 and 0.28, respectively. We can conclude that the best agreement can be found between the 4-clustering results of *k-means* method and the 6-clustering results of *H-ClA* method.

In the next section, we analyze the various clustering results from the point of view of the student answering strategies in order to give meaning to the found partitions.

## IV. Discussion

The interpretation of *ClA* results mainly involves the identification of the typical features characterizing answers of students belonging to the same cluster as well as differences and similarities in answering strategies of students belonging to different clusters.

The four questions in our questionnaire mainly refer to: I) the definition of a physics model, II) the students' beliefs about ways of representing physics models, III) the main characteristics of models, IV) the students' beliefs about the modeling process. We have classified student answers into categories, i. e. the students' answering strategies.

### A. Non-Hierarchical Clustering (NH-ClA)

Looking at *NH-ClA* results, four clusters ($Cl_1, Cl_2, Cl_3, Cl_4$) have been identified. They are characterized by the related centroids and each centroid is represented by one array $\bar{a}_k$, which identifies some answering strategies for each question. These strategies are defined as follows: $\bar{a}_1$ = (1B, 2C, 3B, 4A), $\bar{a}_2$ = (1B, 2B, 3E, 4A), $\bar{a}_3$ = (1C, 2B, 3A, 4A), $\bar{a}_4$ = (1C, 2C, 3B, 4B), where the codes in brackets refer to the questionnaire answer strategies reported in Appendix B. We have already pointed out that the array describing each cluster centroid contains the answers most frequently supplied by the students belonging to the cluster, and in this sense we can identify at what frequency each answering strategy is shared by the cluster students.

In particular, cluster $Cl_4$ is mainly composed of students that use high-level answering strategies to deal with the concepts in the questionnaire. In fact, these students recognize that a model *is a mental representation of a real object or phenomenon, which takes into account the characteristics that are significant for the modeler* (1C). They also think that models *are creations of human thought and their creation comes from continuous interaction with the "real" external world and from its simplification* (2C) and that a model *must highlight the variables that are relevant for the description and/or explanation of the phenomenon analyzed (or the object studied) and their relationships* (3B). *The modeling process is seen as a construction where the model can still contain errors or uncertainty connected with the possibility (or ability) to carefully reproduce the characteristics we are interested in (4B)*. Such students show a conception of model similar to that of expert as defined by Grossslight et al. [33] as general level 3 that is: models as multiple representations, models as construction to test ideas or models as explanatory tools. Such ideas are also described in Treagust et al. paper [34] as student relevant ideas in order to understand the role of scientific models in learning science. Also Pluta et al.[35] identify such model characteristics as good epistemic criteria for scientific models. We cannot perform quantitative comparisons among our results and the ones in the literature for two reasons: from one side on the one hand the differences in the analyzed samples (context, curriculum,....), from the other side on the other hand since data reported in the analyzed references refer to percentages of individual characteristics ( models as multiple representations or models as explanatory tools,....) while our $Cl_4$ students show a model conception that share all these characteristics.

Students in cluster $Cl_2$ show the weakest understanding of the model concept. They refer to a model as *a simple phenomenon or the exemplification of a phenomenon through an experiment or a reduced scale reproduction of an object* (1B), and believe that *models are simple creations of human thought like mathematical formulas, or physics laws and/or they are what we call theories or scientific method* (2B), and give answers regarding the main characteristics of a model that are confused and unclear (3E). For these students *every natural phenomenon can be simplified in order to be referred to a given model* (4A).

$Cl_2$ students can be reported to the level II modelers of the classification scheme developed by Grosslight et al. [33]. Level II modelers see models as representations of real-world objects or events and not as representations of ideas about real-world objects or events, but they realize that there is a specific purpose that guides the way the model is constructed. Similar results have been also obtained in other studies, as for example paper Treagust et al. [34], that groups such conceptions in the theme 2: Scientific models as exact replicas. Also some studies involving teacher conception of scientific models [38, 39] identify the origin of such a realistic conception of scientific model from one side on the students' experiences of everyday models (a scale replica or a precise representation which has accuracy and details), from the other side on teachers' focus on the role of models as examples of objects/processes or their simplifications.

To sum up, we can say that the students in cluster $Cl_4$ seem to share many conceptions connected with an epistemological constructivist view [34]. Students in cluster $Cl_2$, on the other hand, often held beliefs that correspond with a "naïve realist" epistemology [34, 35].

Students in clusters $Cl_1$ and $Cl_3$ do not show a full coherence in their answers, although in different ways. $Cl_1$ students seem to share with $Cl_2$ students the ideas concerning the definition of physics models and the modeling process, but they also share their beliefs about the function as well as the characteristics of physics models with the students from cluster $Cl_4$. In fact, they state that *physics defines models as a simple phenomenon or the exemplification of a phenomenon through an experiment or a reduced scale reproduction of an object (1B)*. However, they also say that they *are creations of human thought and their creation comes from continuous interaction with the "real" external world and from its simplification* (2C). Furthermore, they seem to share the idea that in a modeling process it is important *to highlight the variables that are relevant for the description and/or explanation of the phenomenon being analyzed (or the subject being studied) and their relationships* (3B) and *that every natural phenomenon can be simplified in order to be referred to a given model* (4A).

According to literature that analyzes student modeling in different physics fields [11, 12, 37, 40] we can infer that students belonging to this cluster share a ''hybrid'' [12] or ''synthetic'' [37] conception of scientific model by referring to composite conceptions that unify different features of naïve conceptions and scientifically accepted ones.

Students in cluster $Cl_3$ share the idea that a model is a *mental representation aimed at describing a real object or a phenomenon, which takes into account the characteristics that are significant for the modeler* (1C). However, they also think that *models are simple creations of human thought, like mathematical formulas or physics laws, and/or they are what we call theories or scientific method* (2B). These ideas are not completely consistent with the characteristics assigned to the model or with the students' ideas about the modeling process. In fact they declare that a model *must contain all the rules or all the laws for a simplified description of reality and/or it must account for all the features of reality* (3A) and *that every natural phenomenon can be simplified in order to be referred to a given model* (4A). Their focus on the process of *"simplification"* is also made explicit in the examples they report in order the explain their sentences. For example, many of such students agree that motion without friction is a model, as well as the ideal gas, but do not consider motion with friction or the real gases as models, and explicitly mark them only as really existing situations.

On the other hand, it must be taken into account that the value of the reliability, $r_k^{norm}$, of the $C_3$ centroid is the lowest, showing that the array $\bar{a}_3$ is not very significant in representing the answering strategies of the cluster students. Also, looking in detail at the $\bar{a}_3$ array, the answering strategies are not easily understandable from the point of view of consistency and although they represent the answers most commonly given by $Cl_3$ students, these do not have very high frequencies. For example, no more than 38% is assigned to category 1C. Other answers were also given by a large number of students; for example answering strategy 1B (*A physics model is a simple phenomenon or the exemplification of a phenomenon through an experiment or a reduced scale reproduction of an object*) was selected by 30% of $Cl_3$ students. In our opinion, this may show that a substructure is present in cluster $Cl_3$, and this can be analyzed through results of *H-ClA*, which points out a higher number of clusters.

Moreover, it is noteworthy that *NH-ClA* allows us to quantitatively express the different behavior of students in the different clusters by means of a distance parameter which supplies the distances between the cluster centroids. Looking at the distances between couples of centroids in Fig. 4 we can easily identify that $C_4$ and $C_2$ are the most far apart, and this corresponds to the maximum difference in the behaviors of students belonging to such clusters.

*Hierarchical Clustering (H-ClA)*

The six clusters obtained by *H-ClA* are characterized by the answer strategies reported in Table III. It shows that clusters $\alpha$, $\beta$ and $\gamma$ are closely related to clusters $Cl_4$, $Cl_1$ and $Cl_2$ (obtained through *NH-ClA*), respectively. Clusters $\alpha$, $\beta$ and $\gamma$, although containing a slightly different number of students, include the majority of students of clusters $Cl_4$, $Cl_1$ and $Cl_2$, respectively. Cluster $\alpha$ includes more than half of $Cl_4$ students who mostly exhibit the same characteristics of $Cl_4$ centroid. Cluster $\beta$ includes all students previously grouped in cluster $Cl_1$ and the few added students have not altered the cluster characteristics. Cluster $\gamma$ includes ten students previously included in $Cl_3$ and all the ones previously grouped in cluster $Cl_2$. The answer sheets of all these students that have changed their placing ( from cluster $Cl_3$ to 3 different clusters) were individually analyzed and their position in cluster $Cl_3$ was identified. Almost all were located on the border between different clusters and this fact may explain the new clustering of *H-ClA*.

Clusters $\delta$, $\varepsilon$ and $\zeta$ contain about 80% of students from cluster $Cl_3$ of *NH-ClA*, which are now divided into three groups. Students in cluster $\delta$ give a definition of what a model is analogous to that of $Cl_4$ students, but they are more focused on the concept of a physics model as mathematical model (*models are simple creations of human thought like mathematical formulas, or physics laws and/or they are what we call theories or scientific method* (2B)), and on the characteristics of models, *like simplicity and/or uniqueness and/or comprehensibility* (3D). In the modeling process, they seem to give priority to the process of simplification (*every natural phenomenon can be simplified in order to be referred to a given model* (4A)).

Cluster $\varepsilon$ groups students with a good understanding of the concept of what a model is. This group is the only one that includes in the model definition the function of making predictions (*a model is a simplified representation describing a phenomenon aimed at the understanding of its mechanisms of functioning (or at explaining it or at making predictions (1D))*. Moreover, they describe the representational mode of models as *creations of human thought and their creation comes from continuous interaction with the "real" external world and from its simplification* (2C). Similar ideas are also involved in their definitions of model characteristics (*a model must contain all the rules or all the laws for a simplified description of reality and/or it must account for the features of reality*) (3A). The same can be said for their description of the modeling process (*a model can still contain errors or uncertainty connected with the possibility (or ability) to carefully reproduce the characteristics we are interested in* (4B)).

Cluster $\zeta$ groups students with a weak understanding of the model concept. Models are described as a *set of variables, rules, laws, experiments or observations that simplify reality and represent it on a reduced scale* (1A). Students in this cluster think that *models really exist and are simple, real life situations or simple experiments and humans try to understand them, sometimes only imperfectly* (2A). Among the models' characteristics they focus *on all the rules or all the laws for a simplified description of reality and/or it must account for all the features of reality* (3A). Consistently with these ideas, they do not think that *all natural phenomena can be modeled. There are phenomena that still have not been explained, but perhaps they will be in the future* (4D).

**B. Answers to our research questions**

The analysis of answer strategies elicited by our student sample allows us to answer our first research question: RQ1 - *How are the two different ClA methods effective in partitioning students into groups that can be characterized by common traits in students' answers and how can the results be combined to create a coherent characterization of the data?*

Our results show that our *ClA* methods produce partitions of the student sample into groups that are characterized by common trends in questionnaire answers. However, some of such groups are clearly differentiated for their conceptions about the nature, characteristics and function of physics model, while other groups show conceptions only partially differentiated. Moreover, the two methods show a different "sensitivity" in the clustering procedure. In fact, two clusters ($Cl_1$ and $Cl_2$), resulting from *NH-ClA* methods, almost completely maintain their individuality in *H-ClA*. The other two ($Cl_3$ and $Cl_4$) undergo a redistribution of their elements in a larger number of clusters. This was expected on the basis of the parameters characterizing such clusters, i. e. the spreading of $Cl_4$ and the reliability of $Cl_3$ centroid. *H-ClA* method reassigns some border line students of $Cl_4$ to other cluster and distribute the $Cl_3$ students to three different and smaller clusters. These students, nonetheless, show consistent answering strategies from the point of view of an expert.

We are aware that these results depend on the characteristics of our sample, our questionnaire and our initial empirical analysis. Moreover, although there is no way to determine whether one cluster analysis method is more accurate than another, we have shown that *H-ClA* supplied more details than *NH-ClA*.

To address the second research question (RQ2 - *How do the common traits in students' concept of scientific model identified by ClA method relate to literature findings and what new insights do they supply?*), we compared and contrasted the answering strategies of students belonging to different clusters with similar studies involving the scientific model concepts held by students [11, 33, 34, 35] and teachers [36, 37, 38, 39] reported in the literature. We can conclude that the results of cluster analysis agree with the results obtained by more common research methodologies. In fact, many of the response patterns showed by the groups of students identified by *ClA* methods bear remarkable similarity to those previously reported in the literature. In particular, we have identified in our data the conceptions characterizing the three general levels of thinking about models described in paper [33], as well as answers characterized by the five factors/themes identified in paper [34] and the epistemic criteria for good scientific models [35].

However, our analysis of answering strategies elicited by students grouped in different clusters provides more detailed information than those reported in literature. In fact, in each cluster we found that the meaning assigned by each student to the term model is related to its main characteristics, functions, as well as to procedures for model construction. For example, in our results it is not relevant how many students think models as "*set of variables, rules, laws, experiments or observations that simplify reality and represent it on a reduced scale* (1A)" but, it is relevant how many, among such students, also think that models are *real life situations* (cat. 2A) with particular characteristics (cat 3A) and, for these reasons, not all natural phenomena can be modeled (cat. 4D) (cluster ζ). Moreover, data show that sometimes students give the same definition of model, but they differ for the other kinds of answers. This is mainly evident in two cases: i) students belonging to clusters $\beta$ and $\gamma$ ( that see models as real objects or events) supply the same declaratory definition of scientific model, but their interpretation of model functions is different since in cluster $\beta$ students see as relevant the problem of simplification of reality, whereas in the other one the problem of mathematization is considered more relevant; ii) students belonging to clusters $\alpha$ and $\delta$ supply the same declaratory definition of scientific model (model as mental representation), but the first group of students see such representation as aimed at the understanding of real system behaviors, whereas the second one focus on simplicity or uniqueness of models useful for the description of reality. In this second case, it is evident that a different meaning is given to the word representation.

We also have shown that comparisons of percentages of students of our sample sharing a kind of model conception with percentages previously reported in literature are not significant for two reasons: from one side, since samples are taken from different population and also since our *ClA* results highlight relatioships among different reasoning strategies used by students and analyzed as a whole.

Moreover, the *k-means* results allow us to quantify how the four clusters we identified are different and this gives insights about how different the students' conceptions are. For example, the distance between clusters $C_2$ and $C_4$ reported in Figure 4 is the highest of all distances between couples of clusters, and this is reflected in the categories expressed by the respective centroids, that are most different. In fact, these centroids represent completely different conceptions held by the students contained in these clusters with respect to characteristics and functions of scientific models, as well as to the modeling procedures.

As final remark about the usefulness of applying *ClA* methods, we would like to outline that *ClA* can separate students into groups that can be characterized by common traits in their answers without any prior knowledge on the part of the researcher of what form those groups. This form of bias is more evident in traditional methods of qualitative analysis as well as in some methods of quantitative analysis. However, the researchers must address the problem of how to adjust the problem of different groups supplied by different *ClA* methods.

## V. Conclusions

In this paper, we discussed the problem of quantifying data in order to analyze how to identify groups with common behavior, ideas, beliefs and conceptual understanding in a sample of students. We presented two methods of cluster analysis (*NH-ClA* and *H-ClA*) and analyzed definitions, variables and algorithms in detail, in order to understand the possibilities offered by such methods and their limits. We gave an example of their application in order to demonstrate the necessary approximations and the different ways of interpreting results. The example is an analysis of the answers to a questionnaire given to a sample of university students. The results of this analysis indicate that the two methods are consistent, even if not completely, and *H-ClA* supplies a more detailed partition of our sample into clusters. We tried to interpret the discrepancies through the interpretation of answering strategies of students from different clusters.

It is well known that there is no way to decide whether one clustering method is more significant than another one [30]. The relevance of each clustering method is related to the research content. However, we think that in *H-ClA* the calculations of the consistence/inconsistence of each link (i. e. how relevant the link is in relation to other links in the same hierarchical order) can provide the necessary instrument to analyze clusters in a more detailed way. We have also outlined what subjective choices are at the basis of each algorithm. For example, in the case of *H-ClA* we had to choose between *average and weighted average linkage* and we chose the second one, on the basis of the cophenetic correlation coefficient values. However, we verified the evolution of cluster analysis by choosing the average linkage at the beginning, and obtaining results that were not qualitatively different. We chose the weighted average linkage since in this case the results were more consistent with those of *NH-ClA*. Furthermore, it is well known that the results of *ClA* are only valuable if researchers are able to give meaning to them [6,43], and we have found that our choice made it easier for us to make an interpretation.

It is worth remembering that data that are quantitatively analyzed are often the result of an empirical categorization of raw data (the individual student answers) and this reduction of the initial data can be subject to errors, which obviously influences the final evaluation and the inference about the reasoning strategies supporting students' answers. Such errors can only be reduced (through a clear process of coding and subsequent categorization) and not eliminated, and this must be taken into account when we try to infer typical students' reasoning strategies.

Looking at the meaning of the concept of physics model as it is understood by the students in our sample, our results are consistent with those described in the literature, which illustrate a continuum of ideas/beliefs ranging from naive conceptions to constructivist ones. Our analysis gives details of student conceptions about the function of a physics model and its properties, by identifying features of intermediate conceptions as well as groups of students sharing such conceptions, in a continuum of this type. Moreover, we have been able to quantitatively express the different behavior of students in the different clusters by means of a distance parameter related to the correlation among the student answers.

In conclusion, the quantitative results, obtained with the two ClA methods, have provided us with more detailed information than that reported in the literature, since, in each cluster, the meaning assigned by each student to the term model is related to its main characteristics, functions as well as to procedures for model construction. The characteristics of student clusters, previously described, allow us to draw this general conclusion: in addition to the group of students who exhibit a conception of scientific model basically consistent with a realistic epistemology, the majority of students use the term "representation" (of objects, events, reality) in their definition of scientific model, yet such a term seems used with different meanings. Some intend representation as merely "the way that something is shown and/or described", others go beyond this conception since include the need to define the model constituents and their behaviors in order to make a representation able to explain the object/event that is represented. Only very few, that see the representation as a way to refer also to a conjectured system, are able to identify the predictive function as relevant among different functions of scientific models.

## ACKNOWLEDGMENTS

We are indebted to Professor Rosa Maria Sperandeo-Mineo for her continuous advice and support during the development of this study.

# APPENDIX A

# Mathematical Details

## A.1 The Silhouette Function

When the *k-means* clustering method is applied, in order to choose the number of clusters, $q$, to be initially used to perform the calculations, the so-called *Silhouette Function, S,* [20,21] is defined. This function allows us to decide if the partition of our sample in $q$ clusters is adequate, how dense a cluster is, and how well it is differentiated from the other ones.

For each selected number of clusters, $q$, and for each sample student, $i$, assigned to a cluster, $k$ with $k=1, 2,..q$, a value of the *Silhouette Function*, $S_i(q)$, is calculated as

$$S_i(q) = \frac{\min_k \left[ \sum_{l=1}^{N-n_k} \frac{d_{il}}{N-n_k} \right] - \sum_{j=1}^{n_k} \frac{d_{ij}}{n_k}}{\max \left[ \sum_{j=1}^{n_k} \frac{d_{ij}}{n_k}, \min_k \left[ \sum_{l=1}^{N-n_k} \frac{d_{il}}{N-n_k} \right] \right]}$$

where the first term of the numerator is the minimum average distance of the $i$-th student to students *in different clusters than the one, k, where he/she is placed*, minimized over clusters. The second term is the average distance between the $i$-th student and the other students *in the same cluster k*.

$S_i(q)$ gives a measure of how similar student $i$ is to the other students in its own cluster, when compared to students in other clusters. It ranges from -1 to +1: a value near +1 indicates that student $i$ is well-matched to its own cluster, and poorly-matched to neighboring clusters. If most students have a high silhouette value, then the clustering solution is appropriate. If many students have a low or negative silhouette value, then the clustering solution could have either too many or too few clusters(i.e. the chosen number, $q$, of clusters should be modified).

Subsequently, the values $S_i(q)$ can be averaged on each cluster, $k$, to find the average silhouette value in the cluster, $\langle S(q) \rangle_k$, and on the whole sample, to find the total average silhouette value, $\langle S(q) \rangle$ for the chosen clustering solution. Large values of $\langle S(q) \rangle_k$ are to be related to the cluster elements being tightly arranged in the cluster $k$, and vice versa [20]. Similarly, large values of $\langle S(q) \rangle$ are to be related to well defined clusters [20]. It is, therefore, possible to perform several repetitions of the cluster calculations (with different values of $q$) and to choose the number of clusters, $q$, that gives the maximum value of $\langle S(q) \rangle$.

It has been shown [44] that for values of $\langle S(q) \rangle < 0.50$ reasonable cluster structures cannot be identified in data. If $0.51 < \langle S(q) \rangle < 0.70$ the data set can be reasonably partitioned in clusters. Values of $\langle S(q) \rangle$ greater than 0.70 show a strong cluster structure of data. Figure A.1 shows a partition of 150 data in two (Fig. A.1a), three (Fig. A.1b) and four (Fig. A.1c) clusters. It is easy to see that in all the three cases a partition in clusters is not easily found and this is confirmed by the low values of $\langle S(q) \rangle$ in each of the three partition attempts.

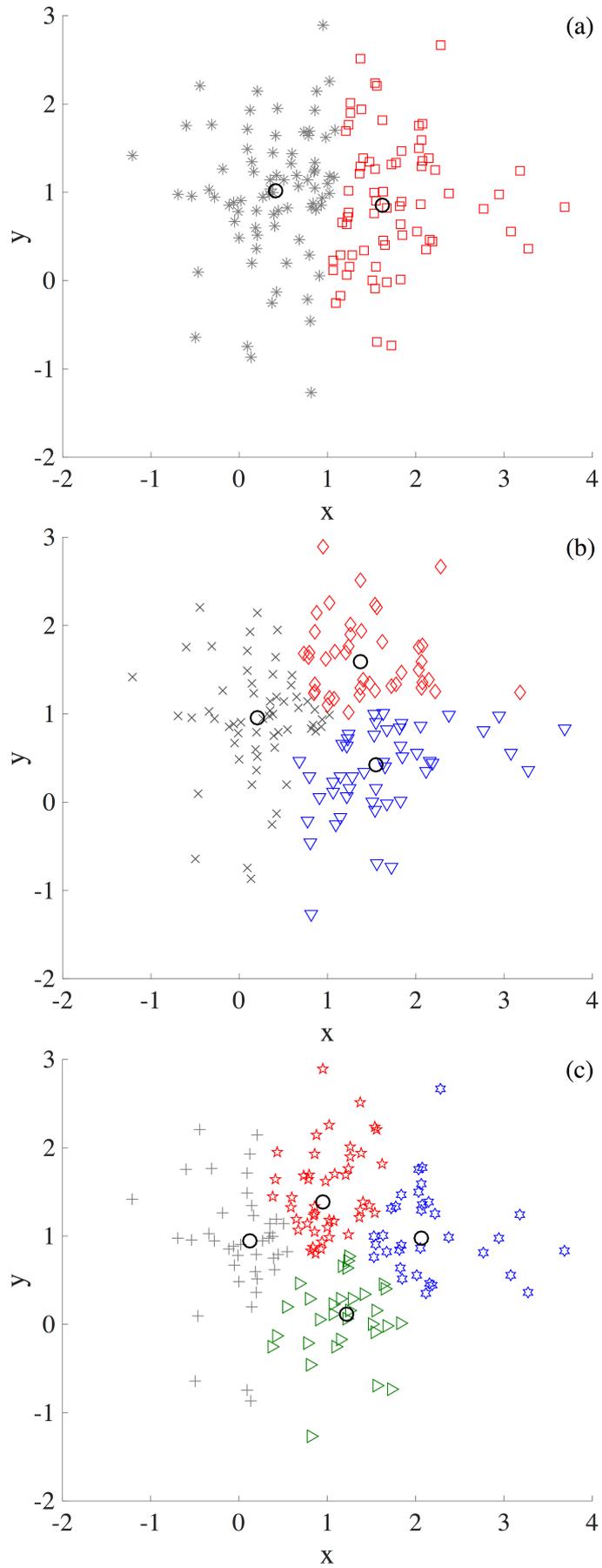

Figure A1. A set of 150 data partitioned in two (a), three (b) and four (c) clusters. The mean values of the Silhouette function are 0.47, 0.45 and 0.45, respectively.

## A.2. An iterative procedure to find the array describing a cluster centroid

In order to find an array that describes the centroid of a cluster by starting from the coordinates of the centroid in the 2-dimensional Cartesian Plane reporting the results of a k-means analysis, we devised a method that consists of repeating the *k-means* procedure in reverse, by using the iterative method described as follows. For each cluster, $Cl_k$, we define a random array $\bar{a}'_k$ (composed of values 1 and 0, randomly distributed) and we calculate the following value

$$\sigma = \sum_i |d_{ik} - d'_{ik}|$$

where $d'_{ik}$ is the distance between the random array and the student, *i*, (belonging to the same cluster $Cl_k$) and $d_{ik}$ is the distance between the centroid and the same student.

By using an iterative procedure that permutes the values of the random array $\bar{a}'_k$, we minimize the σ value and we find the closest array representation[9], $\bar{a}_k$, of the real centroid of $C_k$.

## A.3. Linkage methods

Table A_1 reports the recurrence relationships applied in order to calculate the ultrametric distances, δ, for the different linkage methods, on the basis of the Euclidean distances, $d_{ij}$, represented in matrix, D, (see Section II.A)

Suppose *r, p* and *q* are existing clusters and cluster *r* is the cluster formed by merging *p* and *q* (*r = p ∪ q*). The distances between the elements of *r* and the elements of another cluster *s* are defined for the four linkage methods, as shown in Table A_I [42], where $n_r$ indicates the number of students in cluster *r*, $n_s$ indicates the number of students in cluster *s*, $x_{ri}$ is the *i*-th student in *r* and $x_{sj}$ is the *j*-th student in *s*.

TableA_1
"Ultrametric" distance formulas of commonly used linkages

| | |
|---|---|
| Single linkage | $\delta(r,s) = \min\{d(x_{ri}, x_{sj})\}\ i \in (1,...n_r), j \in (1,...n_s)$ |
| Complete linkage | $\delta(r,s) = \max\{d(x_{ri}, x_{sj})\}\ i \in (1,...n_r), j \in (1,...n_s)$ |
| Average linkage | $\delta(r,s) = \frac{1}{n_r n_s}\sum_i \sum_j d(x_{r_i}, x_{s_j})$ |
| Weighted average linkage | $\delta(r,s) = \frac{\delta(p,s) + \delta(q,s)}{2}$ |

*Single Linkage* links the two clusters *r* and s by using the smallest distance between the students in *r* and those in *s*; *Complete Linkage* uses the largest distance between the students in *r* and the ones in *s*; *Average Linkage* uses the average distance between the students in the two clusters; *Weighted Average Linkage* uses a recursive definition for the distance between two clusters. If cluster *r* was created by combining clusters *p* and *q*, the distance between *r* and another cluster *s* is defined as the average of the distance between *p* and *s* and the distance between *q* and *s*.

As we said in Section II.C, the difference between dendrograms obtained by using the average and the weighted average methods are evident only when the number of elements is not too low. Here, we report an example for a sample of 7 elements. Table A_II supplies the matrix of distances between the 7 elements and Figures A.2 a and b show the two dendrograms for the average and weighted average linkage, respectively.

---

[9] As usual in a procedure to minimize an objective function (in our case, σ), the result may not be unique. In order to be sure to obtain an absolute minimum of (3) we repeated the procedure several times, each time changing the initial conditions, i.e. array $\bar{a}'_k$.

The figure shows some differences, as for example the values of the highest linkage: δ = 1.08 (a) and δ = 1.2 (b).

Table A_II
Matrix of distances of a generic sample of 7 elements

|   | A | B | C | D | E | F | G |
|---|---|---|---|---|---|---|---|
| A | 0 | 0.2 | 0.28 | 0.2 | 0.14 | 0.42 | 1.01 |
| B |   | 0 | 0.2 | 0.28 | 0.14 | 0.42 | 1.01 |
| C |   |   | 0 | 0.2 | 0.14 | 0.42 | 1.01 |
| D |   |   |   | 0 | 0.14 | 0.42 | 1.01 |
| E |   |   |   |   | 0 | 0.4 | 1 |
| F |   |   |   |   |   | 0 | 1.4 |
| G |   |   |   |   |   |   | 0 |

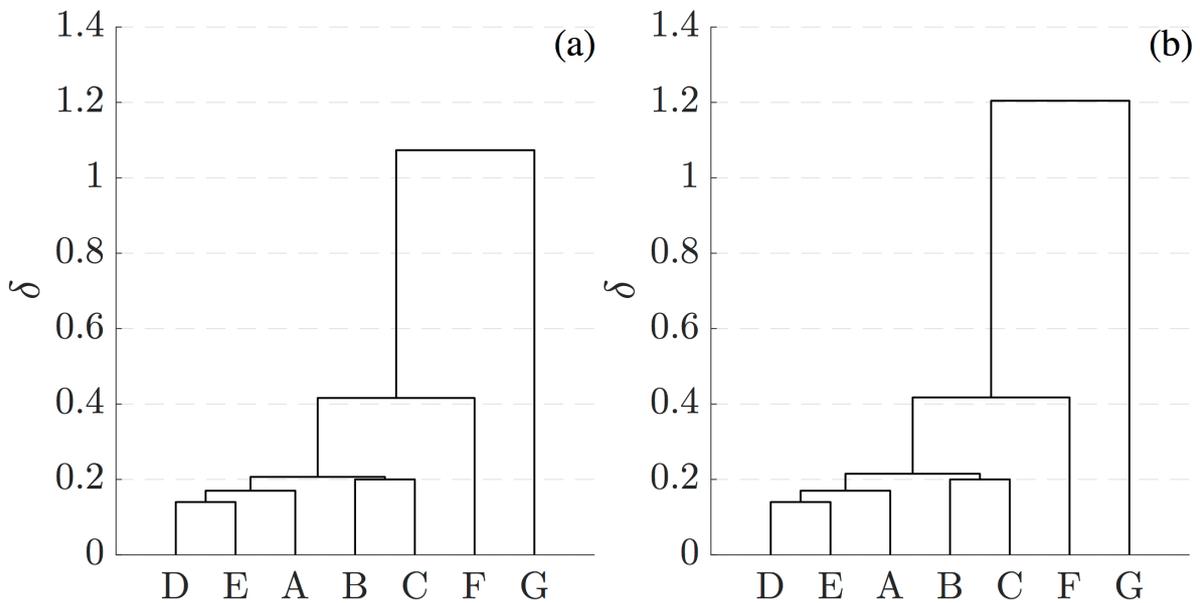

Figure A.2: (a) dendrogram obtained from an average linkage(a) and weighted average linkage (b) of the 7 element sample whose distances are reported in table A_II.

### A.4. Definition of cophenetic correlation coefficient.

The cophenetic correlation coefficient, $c_{coph}$, gives a measure of the concordance between the two matrixes: matrix $D$ of the distances and matrix $\Delta$ of the ultrametric distances. It is defined as

$$c_{coph} = \frac{\sum_{i<j}(d_{ij} - <D>)(\delta_{ij} - <\Delta>)}{\sqrt{\sum_{i<j}(d_{ij} - <D>)^2 \sum_{i<j}(\delta_{ij} - <\Delta>)^2}}$$

where:

- $d_{ij}$ is the distance between elements $i$ and $j$ in $D$.
- $\delta_{ij}$ is the ultrametric distance between elements $i$ and $j$ in $\Delta$, i. e. the height of the link at which the two elements $i$ and $j$ are first joined together.
- $<D>$ and $<\Delta>$ are the average of $D$ and $\Delta$, respectively.

High values of $c_{coph}$ indicate how much the matrix $\Delta$ is actually representative of matrix $D$ and, consequently, how much ultrametric distances, $\delta_{ij}$, are representative of distances, $d_{ij}$.

**A.5 Definition of inconsistency coefficient**
The inconsistency coefficient compares the height of each link in a cluster tree made of $N$ elements, with the heights of neighboring links above it in the tree.
The calculations of inconsistency coefficients are performed on the matrix of the ultrametric distances, $\Delta$, generated by the chosen linkage method.
We consider two clusters, $s$ and $t$, whose distance value is reported in matrix $\Delta$, and that converge in a new link, $k$, (with $k= 1, 2, \ldots N$-1). If we indicate with $\delta(k)$ the height in the dendrogram of such a link, its *inconsistency coefficient* is calculated as follows

$$I_k = \frac{\delta(k) - \langle \delta(k) \rangle_n}{\sigma_n(\delta(k))}$$

Where $\delta(k)$ is the heights of the link $k$, $\langle \delta(k) \rangle_n$ is the mean of the heights of $n$ links below the link $k$ (usually $n = 3$ links are taken into account), and $\sigma_n(\delta(k))$ is the standard deviation of the heights of such $n$ links.

This formula shows that a link whose height differs noticeably from the height of the $n$ links below it indicates that the objects joined at this level in the cluster tree are much farther apart from each other than their $n$ components. Such a link has an high value of $I_k$. On the contrary, if the link, $k$, is approximately the same height as the links below it, no distinct divisions between the objects joined at this level of the hierarchy can be identified. Such a link has a low value of $I_k$.

**A.6 Variation Ratio Criterion (VRC)**
For a partition of $N$ elements in $q$ cluster, the *VRC* value is defined as:

$$\frac{BGSS}{q-1} / \frac{WGSS}{N-q}$$

where *WGSS (Within Group Squared Sum)* represents the sum of the distance squares between the elements belonging to a same cluster and *BGSS (Between Group Squared Sum)*, defines the sum of the distance squares between elements of a given cluster group and the external ones.

# APPENDIX B

## Questions, typical answering strategies and examples of specific student answers

**Q1. The term "model" is very common in scientific disciplines, but what actually is the meaning of "model" in physics?**

| | |
|---|---|
| 1A) A set of variables or rules or laws or experiments and observations that simplify reality and represent it in a reduced scale. | *A model is a general law or an abstract method, used to represent reality in a reduced way.*<br><br>*A model is a set of conventional rules, experiments and observations aimed to simplify and describe Nature* |
| 1B) A simple phenomenon or the exemplification of a phenomenon through an experiment or a reduced scale reproduction of an object. | *As model we intend a physical phenomenon simplified by means of an experiment.*<br><br>*A model is an object that copy in a small scale a real one.* |
| 1C) A mental representation aimed at describing a real object or a phenomenon, which takes into account the characteristics significant for the modeler. | *A model is an idealization of a phenomenon, that allows the researcher to describe what he thinks about its characteristics.*<br><br>*A model is a product of the researcher mind. It is aimed to describe an object or a phenomenon and its features.* |
| 1D) A simplified representation describing a phenomenon aimed at the understanding of its mechanisms of functioning (or at explaining it or at making prediction). | *A model is an ideal representation of a phenomenon that allows us to explain how the phenomenon works.*<br><br>*A model is a representation of reality based on the scientific method that allows us to explain what happens and also to make predictions.* |
| 1E) No answer or not understandable answer | |

**Q2. Are the models creations of human thought or do they already exist in nature?**

| | |
|---|---|
| 2A) Models really exist and are simple, real life situations or simple experiments and humans try to understand them, sometimes only imperfectly. | *A model is a natural phenomenon that is reproduced in laboratory to be studied.*<br><br>*A model is a simple experiment that we do to reproduce a physical situation and to try to understand it, often only roughly.* |
| 2B) Models are simple creations of human thought like mathematical formulas, or physics laws and/or they are what we call theories or scientific method. | *Models are creations of human mind, expressed in a mathematical form.*<br><br>*A model is part of the scientific method. It is created by human thought and it is resumed in laws and theories.* |
| 2C) Models are creations of human thought and their creation comes from continuous interaction with the ''real'' external world and from its simplification. | *Models are the creation of human thought. They are abstractions coming from the real phenomena in order to simplify them.*<br><br>*A model is the creation of human mind drown from scientists' observation of natural phenomena.* |
| **2D)** Models are creations of human thought aimed at explaining natural phenomena and making predictions. | *A model is an artificial creation of man. It is based on the observation of Nature and is aimed at explaining it.*<br><br>*Models are created by the human mind and are aimed at explaining Nature and making predictions.* |
| **2E)** No answer or not understandable answer | |

**Q3. What are the main characteristics of a physical model?**

| | |
|---|---|
| 3A) It must contain all the rules or all the laws for a simplified description of reality and/or it must account for all the features of reality. | *A physical model is characterized by a mathematical formulation that allows us to completely describe the variables we really observe.*<br><br>*The main characteristics of a model are the laws that simplify the description of reality.* |
| 3B) It must highlight the variables that are relevant for the description and/or explanation of the phenomenon analyzed (or the object studied) and their relationships. | *A model is useful if it puts in evidence the variables relevant to understand the phenomenon.*<br>*A model must set the relationships among the variables that we measured during the observation/experimental phase.* |
| 3C) Their characteristics can classify models as descriptive or explicative or interpretative. | *A model must allow us to describe and explain what's happening in nature.*<br><br>*A model is a way to understand the nature.* |
| 3D) Their main characteristics are simplicity and/or uniqueness and/or comprehensibility. | *A model should mainly be comprehensible, so to be used by everyone.*<br><br>*A model must be clear and unique, so to not give ambiguous answers to our questions.* |
| 3E) No answer or not understandable answer. | |

**Q4. Is it possible to build a model for each natural phenomenon?**

| | |
|---|---|
| 4A) Yes, every natural phenomenon can be simplified in order to be referred to a given model. | *Yes. We can always find a simple model for each natural phenomenon.*<br><br>*Yes, because we can choose a simplification level for the model. So, it is always possible to build a model for a phenomenon.* |
| 4B) Yes, but the model can still contain errors or uncertainty connected with the possibility (or ability) of carefully reproducing the characteristics we are interested. | *Yes, but all depends on the complexity of the phenomenon. In some cases we cannot completely reproduce a phenomenon.*<br><br>*Yes, but we will always be able to only partially reproduce a phenomenon, due to uncertainty in measurements and in their description.* |
| 4C) No. There are phenomena that cannot be described or explained with a model and/or that cannot be defined in terms of precise physical quantities. | *Probably not, because we cannot always find variables for all natural phenomena. An example is biology, where we cannot simplify the functioning of the nucleus of the cell with precise variables, but take all into account.*<br><br>*No, because we cannot know all the variables relevant for the description of the phenomenon* |
| 4D) No. There are phenomena that have not been still explained and these, perhaps, will be in the future. | *No. There are phenomena that are not explained due to our limitations and technology.*<br><br>*No, because in some cases we don't have the right mathematical tools to build a model to the level we want. But in the future we will probably have them.* |
| 4E) No answer or answer not understandable | |